\documentclass[runningheads]{llncs}
\usepackage{graphicx}

\usepackage{tikz}
\usepackage{comment}
\usepackage{amsmath,amssymb} 
\usepackage{color}

\usepackage{booktabs}
\usepackage{pifont}
\usepackage{xcolor}
\newcommand{\xmark}{\ding{55}}
\usepackage{subcaption}
\usepackage{tabularx}
\usepackage{multirow}

\usepackage[accsupp]{axessibility}  

\begin{document}
\pagestyle{headings}
\mainmatter
\def\ECCVSubNumber{4926}  

\title{Fast Online Video Super-Resolution with Deformable Attention Pyramid} 

\author{Dario Fuoli \inst{1} \and
Martin Danelljan \inst{1} \and
Radu Timofte \inst{1,2} \and Luc Van Gool \inst{1,3}}
\authorrunning{D. Fuoli et al.}
%
\institute{Computer Vision Lab, ETH Z\"urich, Switzerland \and
University of W\"urzburg, Germany \and
KU Leuven, Belgium
\\
\email{\{dario.fuoli, martin.danelljan, radu.timofte, vangool\}@vision.ee.ethz.ch}}
\maketitle

\begin{abstract}

Video super-resolution (VSR) has many applications that pose strict causal, real-time, and latency constraints, including video streaming and TV.
We address the VSR problem under these settings, which poses additional important challenges since information from future frames is unavailable. Importantly, designing efficient, yet effective frame alignment and fusion modules remain central problems.
In this work, we propose a recurrent VSR architecture based on a deformable attention pyramid (DAP). Our DAP aligns and integrates information from the recurrent state into the current frame prediction. To circumvent the computational cost of traditional attention-based methods, we only attend to a limited number of spatial locations, which are dynamically predicted by the DAP.
Comprehensive experiments and analysis of the proposed key innovations show the effectiveness of our approach. We significantly reduce processing time and computational complexity in comparison to state-of-the-art methods, while maintaining a high performance. We surpass state-of-the-art method EDVR-M on two standard benchmarks with a speed-up of over $3\times$.

\end{abstract}

\section{Introduction}
\label{sec:introduction}

\begin{figure}[t]
\begin{center}
\includegraphics[width=0.73\textwidth]{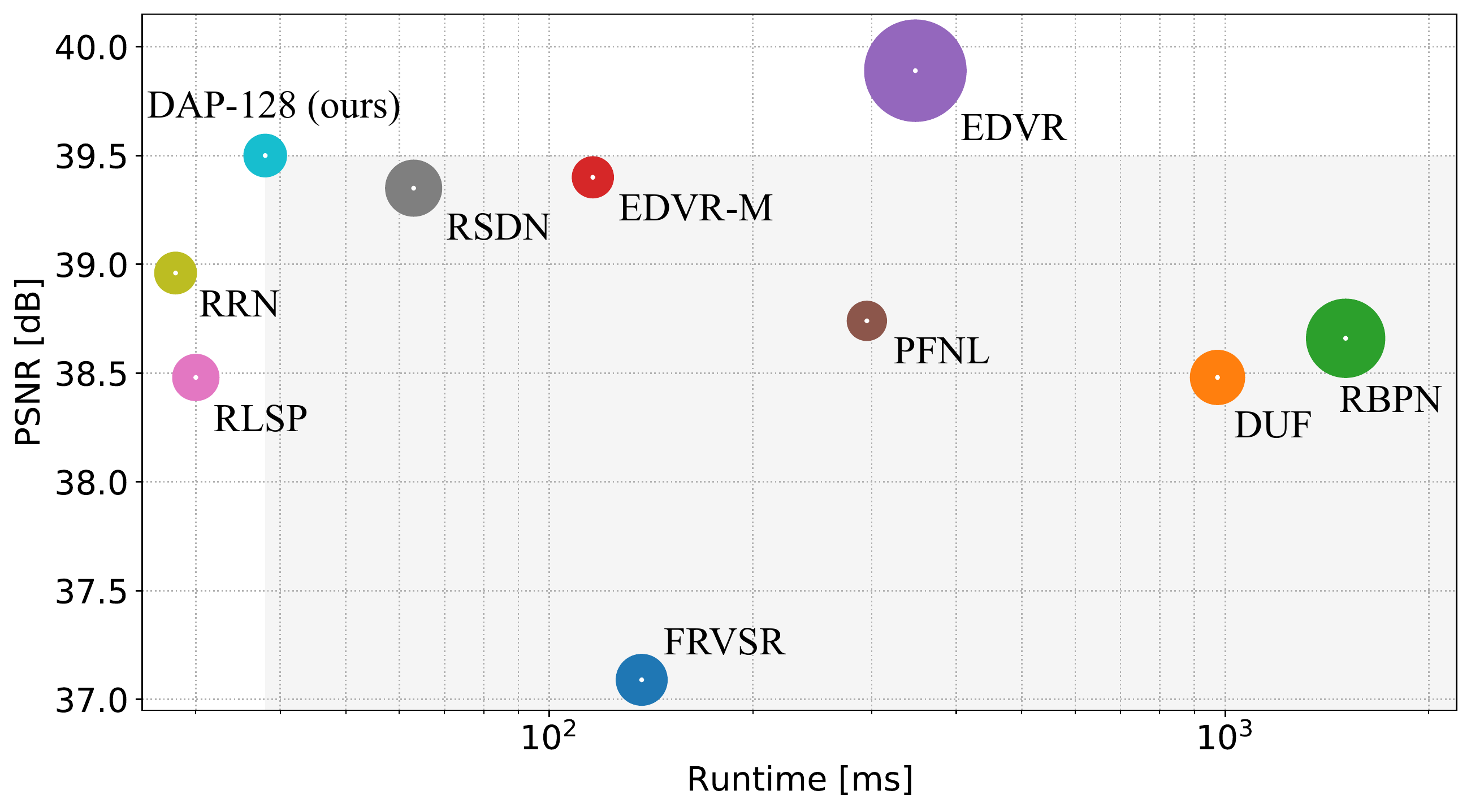}
\end{center}
\vspace{-0.6cm}
   \caption{Runtime vs.\ performance on UDM10~\cite{PFNL}. Disk areas correspond to number of parameters for each method. Our method DAP-128 achieves highly competitive performance with high speed ($38$ms per frame) and minimal complexity. Light gray highlights the Pareto dominant region of DAP-128.}
   \label{fig:teaser}
\end{figure}

Video super-resolution (VSR) is the problem of restoring spatial high-frequency components from low-resolution video frames.
In contrast to single image super-resolution, where methods are bound to rely on image priors, VSR offers the opportunity to utilize additional observations from adjacent frames and long-range temporal correlations to reconstruct a single frame.
For this reason, effective frame alignment and fusion of salient features along the temporal axis constitute the main challenges in VSR.

Many practical applications, including TV and video streaming, depend on the ability to run algorithms online and in real-time, where minimal latency and high-speed processing are essential.
However, hard time constraints in online video processing pose major challenges for learned VSR, as high performance strongly correlates with computational complexity in deep neural networks (DNN). Contrary to many other computer vision problems, it is thus important to carefully optimize the network's performance while minimizing architectural complexity. 
In addition to fast inference, a tailored solution to the problem of online VSR is required. Contrary to many prior works \cite{huang_bidirectional,chan2020basicvsr,chan2021basicvsr++,duf,wang2019edvr,li_mucan,Isobe_2020_CVPR_TGA}, we therefore address the problem of designing a strictly causal VSR approach. This imposes an additional challenge, since causality prohibits the access to information from future frames.

Designing effective yet efficient alignment and fusion methods for VSR brings considerable challenges. Existing methods use inefficient alignment strategies, e.g. expensive alignment in feature space \cite{wang2019edvr}, exhaustive attention computation \cite{PFNL,li_mucan} or ineffective implicit convolution-based alignment \cite{jo2018duf}, without specific care about runtime. Most rely on information from neighboring frames only and neglect the potential of computation reuse between consecutive frames.
As a consequence of missing efficient alignment/fusion mechanisms, fast methods generally avoid such modules entirely \cite{fuoli2019rlsp,isobeRRN,rsdn}.
In this work, we employ a recurrent VSR architecture due to its online nature and efficiency, and address the aforementioned open issues of efficient alignment and fusion.

We propose a VSR approach by taking inspiration from recent advances in attention~\cite{Wang_2018_CVPR_nonlocal} and transformers~\cite{transformer,visualtransformer}. 
As a case in point, attention- and transformer-based solutions have been successfully employed to computer vision tasks.
Attention provides an advantage over convolutions as it allows effective matching and fusion of global information in early layers. Additionally, the operation implicitly serves for alignment to handle the displacements between frames in VSR.
While the mechanism facilitates high performance, its quadratic complexity along with exhaustive correlation computations often render it unsuitable to time critical applications, especially in high-dimensional domains like video. In order to leverage its potential, the attention mechanism requires major adaptations to fulfil the requirements for fast video processing.

We tackle the aforementioned challenges by dynamically predicting pairs to be utilized for attention, thereby averting the high computational complexity in classic attention algorithms.
In particular, we employ a deformable attention pyramid (DAP) for efficient information fusion at dynamically computed locations in the hidden state of our recurrent unit. DAP simultaneously addresses the misalignment and fusion through flexible offset prediction and discriminative aggregation with attention. The deformable attention mechanism allows robust fusion of spatially shifted features and counteracts error accumulation in the hidden state by means of dynamic selection of informative features.

For fast offset prediction we utilize a light-weight convolutional network. However, shallow convolutional networks suffer from small receptive fields due to their locality bias. This drawback limits the ability to handle large spatial displacements caused by movement between frames. We efficiently expand the receptive field by using a pyramid type network comprising a multi-level encoder followed by iterative attention-based offset refinement.
According to the computed offsets, our fusion module effectively aggregates information from the hidden state.
After the alignment/fusion stage, the combined processing of hidden state features and upsampling is performed by residual convolutional blocks, which ultimately output the high-resolution frame and the next hidden state. 

Our experiments show great benefits of our proposed modules to the problem of VSR. An extensive ablation study clearly highlights the effectiveness of our contributions. We significantly reduce processing time and computational complexity in comparison to state-of-the-art methods, while achieving high performance. We attain higher PSNR than state-of-the-art method EDVR-M $+0.06$dB with a speed-up of over $3\times$ on the standard benchmark REDS.

\section{Related Work}

The ability to leverage complementary information in the temporal dimension for improved interpolation quality, represents a major difference between VSR and single-image super-resolution, where restoration algorithms are constrained to rely on priors only.
An overview of recent state-of-the-art VSR methods is provided by \cite{Nah_2019_CVPR_Workshops,Nah2019reds,AIM2019VXSRchallenge,Son_2021_CVPR}.
Two distinct mechanisms have been proposed in the literature to leverage this extra information in VSR; (1) sliding windows \cite{wang2019edvr,PFNL,li_mucan,Isobe_2020_CVPR_TGA,Xiao_2021_CVPR} and (2) recurrent processing~\cite{toflow,frvsr,Haris_2019_CVPR_RBPN,fuoli2019rlsp,rsdn,isobeRRN,chan2020basicvsr,chan2021basicvsr++}. 
Sliding windows extract information from a fixed set of adjacent frames, while recurrent approaches accumulate information over time in a hidden state for exploitation at the current time step. 

\textbf{Window-based}
Earlier methods~\cite{kappeler,tao,caballero,caballeroespcn} compute optical flow (OF) to warp adjacent frames for motion compensation with respect to the center frame.
DUF~\cite{jo2018duf} investigates VSR without explicit motion compensation by applying 3D-convolutions on a set of adjacent frames in combination with dynamic upscaling filters.
Recent window-based designs often achieve higher performance in trade-off with runtime. Such a strategy has the benefit to use extensive parallel processing during training, which facilitates exploration of larger models.
PFNL~\cite{PFNL} adopts non-local residual blocks~\cite{Wang_2018_CVPR_nonlocal} as an alternative to motion estimation in order to progressively fuse information of adjacent frames.
Contrary to other window-based methods, which fuse frames individually, RBPN~\cite{Haris_2019_CVPR_RBPN} introduces a module to iteratively aggregate information from neighboring frames inside a fixed temporal window with recurrent back-projection. 
EDVR~\cite{wang2019edvr} proposes separate modules for alignment and fusion. Frames are aligned in feature space with cascaded deformable convolutions and fused by application of temporal and spatial attention maps.  
MuCAN~\cite{li_mucan} utilizes a hierarchical correspondence aggregation strategy to detect inter-frame correspondences by selecting a fixed set of the most similar patches after an exhaustive search on a local neighborhood. Aggregation from these selected patches is performed by a convolutional block. To adress the issue of misalignment, TGA~\cite{Isobe_2020_CVPR_TGA} splits neighboring frames within a window into groups, according to their temporal distances from the center frame. Fusion is accomplished by application of attention maps.

\textbf{Recurrent}
The temporal receptive field of approach (1) is limited in consequence of its fixed window size and usually depends on the availability of future frames, which introduces latency at inference. Approach (2) has a potentially unlimited temporal receptive field and generally accumulates information more efficiently with reuse of computation through a hidden state. 
Recurrent networks for super-resolution can be further divided into unidirectional~\cite{frvsr,fuoli2019rlsp,rsdn,isobeRRN} and bidirectional methods~\cite{huang_bidirectional,chan2020basicvsr,chan2021basicvsr++}.

\emph{Unidirectional:}
As one of the first, FRVSR~\cite{frvsr} accounts for motion between consecutive frames in a recurrent fashion. The previous high-resolution estimate is warped towards the current frame with OF. Later, RLSP~\cite{fuoli2019rlsp} introduced efficient propagation of implicit information in a hidden state with a fully convolutional recurrent network. The information in the hidden state is accumulated without explicit motion compensation.
RSDN~\cite{rsdn} further improves that concept by dividing the content into structure and detail components. Additionally, it accounts for error accumulation in the hidden state by detecting large displacements between frames.

\emph{Bidirectional:}
In order to leverage long-distance temporal correlations, an aggregation strategy considering information from all frames in a video is favorable and can be efficiently leveraged with forward and backward passes.
While this approach is best suited for high performance, it violates causality -- a necessary property for online inference.
BasicVSR~\cite{chan2020basicvsr} achieves high performance with light recurrent cells by employing two passes over an entire video. Its successor BasicVSR++~\cite{chan2021basicvsr++} further improves performance with extensive bidirectional propagation strategies.
Unfortunately, these approaches can not be evaluated online as they violate causality.
Therefore, we design a unidirectional recurrent network for fast online inference and instead maximize information accumulation from the hidden state with our proposed efficient dynamic module.

Dynamic attention/transformer mechanisms are also explored in unrelated domains like object detection in images~\cite{zhu2020deformable}, video object segmentation~\cite{seong2021hierarchical} and to increase network capacity~\cite{Chen_2020_CVPR}. \cite{zhu2020deformable} proposes a transformer network that attends to dynamically predicted locations in order to detect relevant object features within a single image. Contrary, our proposed DAP leverages attention to efficiently match locations between consecutive frames. Additionally, our DAP leverages the discriminative feature of attention for more robust aggregation/fusion from the hidden state, where information is dynamically merged from multiple locations according to their relevance.
\cite{seong2021hierarchical} employs a top-k memory matching scheme to reduce computational overhead of its attention-based module for video object segmentation.
DCN~\cite{Chen_2020_CVPR} uses dynamic ensemble learning over convolution kernels to increase network capacity. Our DAP is fundamentally different, instead predicting and attending to multiple spatial offset locations to explicitly match and fuse information across large spatial distances.

\section{Method}

\begin{figure}[t]
\begin{center}
\includegraphics[width=0.65\linewidth]{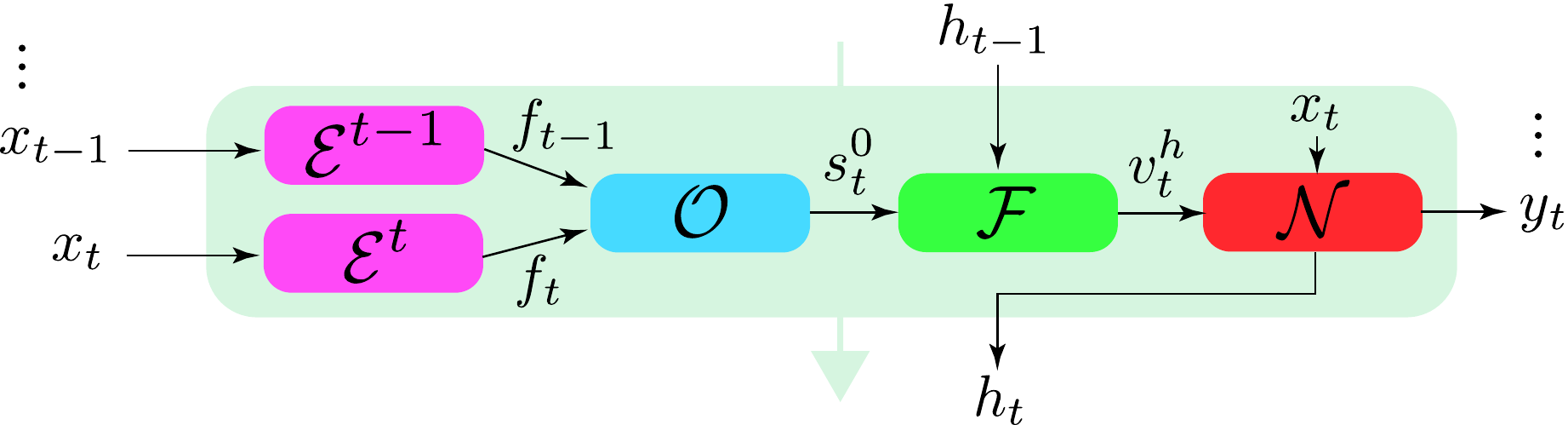}
\end{center}
\vspace{-0.6cm}
   \caption{Schematic overview of our proposed method showing the interactions of our main modules in the recurrent cell.}
\label{fig:overview}
\end{figure}

\subsection{Overview}
According to Nyquist-Shannon's sampling theorem, the frequency band of a discrete signal is band-limited at a specific frequency $f_N$ in the spectrum, called the Nyquist frequency.
A VSR algorithm's task, is to recover the high-frequency content above said frequency from a low-resolution video $x\in\mathbb{R}^{T\times H\times W\times C}$, which is lost after subsampling its high-resolution counterpart $y\in\mathbb{R}^{T\times rH\times rW\times C}$ with scaling factor $r$. 
To fulfil the requirements for online VSR, an effective and efficient algorithm is necessary.

We propose a recurrent algorithm to address the two most important aspects of online VSR with strong emphasis on fast runtimes. Namely, the handling of misalignment between frames in combination with the update/extraction of information in the hidden state $h_t$. Such a setup allows efficient temporal aggregation of salient features from past frames $x_{0:t-1}$.

Recent advances in attention- and transformer architectures led to large performance gains in a wide range of computer vision tasks. 
However, due to pixel-dense processing requirements in VSR, a naive implementation of attention- or transformer type components is highly inefficient and prohibits the application of such operations due to their quadratic computational complexity.
To alleviate this issue, we design an attention mechanism to dynamically predict a subset of relevant key/value pairs in the hidden state, omitting an exhaustive and expensive search over all possible pairs.

In particular, a recurrent cell propagates a pixel-dense hidden state $h_t$. Hidden state fusion and misalignment are simultaneously handled by our proposed deformable attention mechanism DAP.
DAP uses a pyramid type network for dynamic offset prediction. First, our encoder network $\mathcal{E}$, individually divides frames $x_t, x_{t-1}$ into multi-level feature maps $f_t, f_{t-1}$ representing fine-to-coarse views on the input, effectively enlarging the receptive field and enriching representational power. 
\begin{equation}
\begin{aligned}
    f_t, f_{t-1} &= \mathcal{E}^t(x_t), \mathcal{E}^{t-1}(x_{t-1}) \\
\end{aligned}
\end{equation}

Our deformable attention module $\mathcal{O}$ iteratively refines the calculated offsets $s_t$ from coarse to fine. 
\begin{equation}
\begin{aligned}
    s^0_t &= \mathcal{O}(f_t, f_{t-1})  \\
\end{aligned}
\end{equation}
Our fusion module $\mathcal{F}$ then aggregates the hidden state features according to the final offsets. 
\begin{equation}
\begin{aligned}
    v^h_t &= \mathcal{F}(h_{t-1}, s^0_t)    \\     
\end{aligned}
\end{equation}
After the fusion/alignment stage, our main processing unit $\mathcal{N}$, consisting of repeated residual information multi-distillation blocks \cite{Hui-IMDN-2019}, estimates the high-resolution frame $y_t$ and the next hidden state $h_t$.
\begin{equation}
\begin{aligned}
    \left[y_t, h_t\right] &= \mathcal{N}(x_t, v^h_t), \quad t=0, ..., T \\
\end{aligned}
\end{equation}

A high-level overview of our method is shown in Fig.~\ref{fig:overview}, which depicts the recurrent cell and the interactions of its main modules. 
Next, we will explain our proposed modules in more detail.

\subsection{Deformable Attention Pyramid (DAP)} 
In order to fuse the accumulated past information from the hidden state $h_{t-1}$ in relation to the current time step $t$, we employ a deformable attention pyramid. Our module operates on pixel-dense representations to adhere to the low-level processing requirements for VSR.
We design our DAP to aggregate spatially displaced information in a robust and highly efficient manner. To achieve these properties we employ a pyramid type processing module working on encoded multi-level features computed from input frames $x_{t-1}$ and $x_t$ to efficiently enlarge the receptive field. Further, to avoid exhaustive correlation computations we restrict our attention module to a small set of key/value pairs at dynamically predicted offset locations in $x_{t-1}$ for cross-attention with the current frame $x_t$.  

First, offsets from $x_{t-1}$ to $x_t$ are computed. The offsets serve two purposes; (1) the handling of misalignment between frames and (2) a drastic minimization of attention weight computations. According to these offsets, the information is fused by cross-attention for exploitation at time step $t$. The final full-resolution offsets allow robust pixel-dense fusion through cross-attention between current frame $x_t$ and hidden state $h_{t-1}$.

\textbf{Multi-level Encoder}
It is essential in VSR to capture offsets across large distances since there can be fast motion in the camera or objects. As a solution to this problem, we use a multi-level encoder to obtain features in multiple resolutions. The coarser level features serve to capture larger motion due to a larger spatial view on the frames.
We encode features for the last and current input frames $x_{t-1}$ and $x_t$ at levels $l=0, ..., L$. To further enrich a frame's representation at each level, we encode it into higher $d$-dimensional features $f^{l}_t$. The smaller resolution representations at higher levels are obtained by repeated convolutional blocks $\mathcal{C}^l$, consisting of 4 convolutions,  with intermediate bilinear downsampling steps between blocks ($\times 2$) which we denote with the operator $D_\downarrow\left(\cdot\right)$, see
Eq.~\ref{eq:encoder}. Even with small $3\times 3$ kernels, such a strategy increases the receptive field exponentially through the repeated downsampling operations. We employ individual processing chains for both input frames $x_{t-1}$ and $x_t$, we set $L=3$.

\begin{equation}
\label{eq:encoder}
\begin{aligned}
f^{l}_t &= D_\downarrow\left(\mathcal{C}^l\left(f^{l-1}_t\right)\right), & f^{0}_t &= \mathcal{C}^0\left(x_t\right), & l= & 0,...,L\\
f^{l}_{t-1} &=  D_\downarrow\left(\mathcal{C}^l(f^{l-1}_{t-1})\right), & f^{0}_{t-1} &= \mathcal{C}^0\left(x_{t-1}\right), & l= & 0,...,L
\end{aligned}
\end{equation}

\begin{figure}[t]
\centering
\begin{center}
\includegraphics[width=0.9\textwidth]{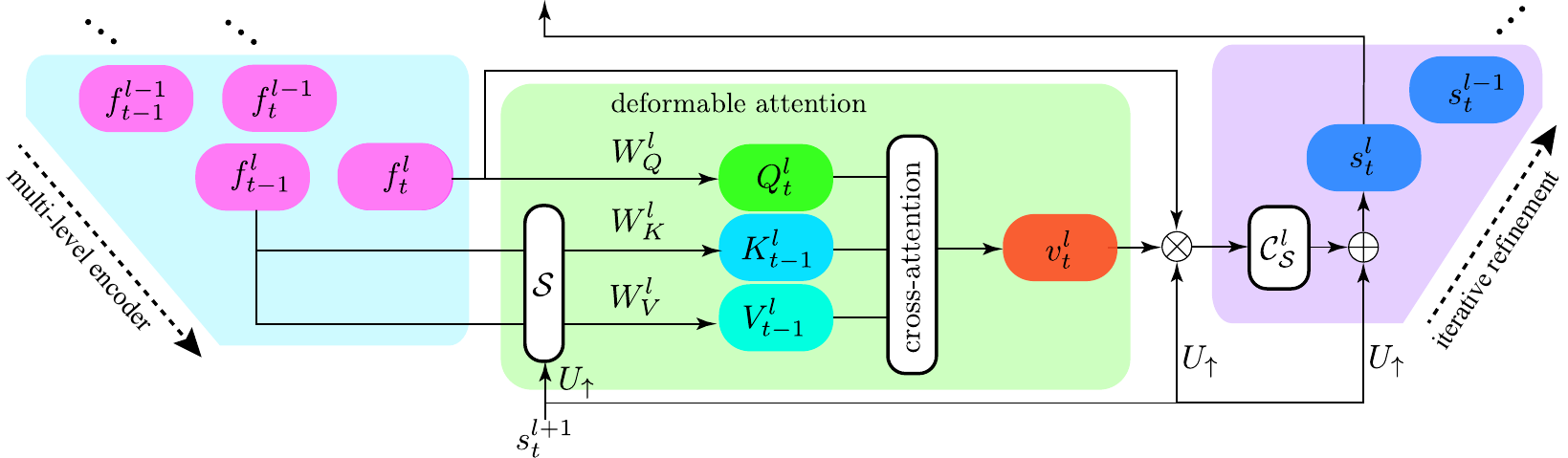}
\end{center}
\vspace{-0.6cm}
   \caption{Deformable attention pyramid (DAP). First we calculate multi-level features from $x_t$, and $x_{t-1}$ using a U-Net~\cite{ronneberger2015unet} type encoder. In each pyramid level $l$, $k$ sampling locations $s^l_t \in \mathbb{R}^{H/2^l\times W/2^l\times 2k}$ are calculated per pixel to serve as key/value locations in the upper level's deformable attention block. Features in the upper level are fused from $t-1$ towards $t$ according to $s^l_t$ with cross-attention, before calculating the residual offsets with convolutional block $\mathcal{C}^l_\mathcal{S}$ with respect to the lower layer $l+1$. Offsets $s^l_t$ are refined iteratively until the obtained locations $s^0_t$ at level $l=0$ are finally employed to perform cross-attention fusion in the hidden state $h_{t-1}$.\\
   $\otimes$: concatenation in channel dimension, $\oplus$: element-wise addition.}
\label{fig:pyramid}
  \vspace{-0.4cm}
\end{figure}

\textbf{Deformable Attention}
In order to significantly reduce complexity of our attention module, we confine the search for salient features to dynamically selected locations in the feature maps, instead of an exhaustive correlation computation over a large neighborhood or even the whole frame. The quadratic component prevalent in attention mechanisms is overcome by applying pure cross-attention towards the current frame $x_t$, resulting in a linear dependency in the number of key/value pairs.
In particular, we process the encoded features $f^{l}_t$, $f^{l}_{t-1}$ with our optimized, light-weight, deformable attention operation mechanism at each pyramid level $l$. We largely reduce the computational effort by merely computing pixel-dense correlations between embedded features representing the queries $Q^{l}_t$ of current frame $x_t$, and key/value pairs $K^{l}_{t-1}$/$V^{l}_{t-1}$ sampled at $k$ dynamically predicted spatial locations $s^l \in \mathbb{R}^{H/2^l\times W/2^l\times 2k}$ in $f_{t-1}$. 
Queries and key/value pairs are linearly embedded with parameters $W^{l}_Q$, $W^{l}_K$ and $W^{l}_V$. We apply scaled dot-product attention with softmax to aggregate the values from $V^{l^T}_{t-1}$.
We account for resolution mismatch between feature maps at level $l$ and sampling locations $s^{l+1}$ from the previous level $l+1$ with bilinear upsampling ($\times 2$), denoted by $U_\uparrow(\cdot)$.
The corresponding equations are presented in Eq.~\ref{eq:attention}, an illustration is provided in Fig.~\ref{fig:pyramid}. Please note the reverse order for pyramid level index $l$, since the processing is performed from coarse to fine.

\begin{equation}
\label{eq:attention}
\begin{aligned}
l &= L,... , 0 \\
Q^{l}_t &= W^{l}_Q f^{l}_t \\
K^{l}_{t-1} &= W^{l}_K \mathcal{S}(f^{l}_{t-1}, U_\uparrow(s^{l+1}_t)) \\
V^{l}_{t-1} &= W^{l}_V \mathcal{S}(f^{l}_{t-1}, U_\uparrow(s^{l+1}_t)) \\
v^{l}_t\left(Q^{l}_t, K^{l}_{t-1}, V^{l}_{t-1} \right) &= \text{softmax}\left(\frac{Q^{l}_t K^{l^T}_{t-1}}{\sqrt{d}}\right)V^{l}_{t-1}
\end{aligned}
\end{equation}

\textbf{Iterative Refinement}
We propose an efficient iterative coarse-to-fine scheme to address blending of multi-level offset representations in different resolutions, attention-aggregated values $v^l_t$, and features $f^l_t$ from the current frame $x_t$. 
In each pyramid level, the dense offsets $s^l_t \in \mathbb{R}^{H/2^l\times W/2^l\times 2k}$ are iteratively refined by adding residual offsets to the previous level's offsets $s^{l+1}_t$ with a light-weight convolutional block $\mathcal{C}^{l}_\mathcal{S}$. Our offset prediction network $\mathcal{C}^{l}_\mathcal{S}$ uses large $7\times 7$ kernels to ensure dense computation with a large receptive field, in contrast to smaller $3\times 3$ kernels employed in all our other modules. 

\begin{equation}
\label{eq:offsets}
\begin{aligned}
l &= L,... , 0 \\
s^{l}_t &= \mathcal{C}^{l}_\mathcal{S}\left(f^{l}_t, v^{l}_t, U_\uparrow(s^{l+1}_t) \right) + U_\uparrow(s^{l+1}_t), \quad s^{L}_t = \mathcal{C}^{L}_\mathcal{S}\left(f^{L}_t\right)
\end{aligned}
\end{equation}

\textbf{Hidden State Fusion}
Ultimately, the top level offsets $s^0_t$ serve to fuse salient hidden state features for exploitation at time step $t$. For that matter, another deformable attention block $v^h_t$ takes care of fusion in full resolution by leveraging the computed offsets $s^0_t$. Since it is critical to minimize the computational effort for fast VSR, our DAP module is processing frames in a lower $d$-dimensional space ($d=8$), because the frames' channels are fixed in size setting a limit on available information. Conversely, a larger channel size in our main processing pipeline -- the size of the hidden state -- increases the upper limit of storable information, which facilitates higher performance. Thus, the hidden state's deformable attention block $v^h_t$ performs query/key matching in $d$-dimensional embeddings, while the values are embedded and aggregated in their native high-dimensional space. We denote our network by DAP-$n$, $n$ represents the feature size in the main processing block.

\begin{equation}
\label{eq:hidden_state_attention}
\begin{aligned}
v^{h}_t\left(Q^{0}_t, K^{h}_{t-1}, V^{h}_{t-1} \right) &= \text{softmax}\left(\frac{Q^{0}_t K^{h^T}_{t-1}}{\sqrt{d}}\right)V^{h}_{t-1}
\end{aligned}
\end{equation}

A significant improvement in runtime is achieved by grouped sampling inside tensors at all stages, which are omitted in the notation for clarity. The number of groups is chosen according to the number of sampled key/value pairs $k=4$.

\section{Experiments}

We conduct comprehensive experiments in our ablation study to highlight our key innovations and compare our best performing configuration to state-of-the-art methods on 3 diverse standard benchmarks REDS~\cite{Nah2019reds}, UDM10~\cite{PFNL} and Vimeo-90K~\cite{xue2019video} with 2 different subsampling kernels (Bicubic and Gaussian). We use the proposed split for REDS according to \cite{wang2019edvr} along with the provided training pairs from \cite{Nah2019reds}. Our results for Vimeo-90K and UDM10 test sets are obtained by application of a Gaussian low-pass filter followed by resampling of every $4$-th pixel along each spatial dimension. Following the literature we set the Gaussian filter's standard deviation and kernel size to $\sigma=1.6$ and $13$ respectively, Vimeo-90K serves as training set for both test sets. 
During training we uniformly crop sequences with spatial size $256\times 256$ (high resolution) and adopt random flips, rotations and temporal inversion to augment the data. The initial learning rate is set to $10^{-4}$ and is reduced after reaching a plateau in two steps to $0.5\times10^{-4}$, then $10^{-5}$. To stabilize training we use gradient clipping. More specific details will be explained in the respective subsections. Similar to recent proposals in the literature our networks are trained with a smooth version of L1 loss \footnote{PyTorch's torch.nn.SmoothL1Loss() with $\beta=10^{-2}$}, which showed benefits over L2 loss for super-resolution as it is less sensitive to outliers. Our model can be trained end-to-end without relying on pretrained modules or external data. We use Adam optimizer~\cite{adam} and set the scaling factor to $r=4$ in all our experiments.

\subsection{Ablation}
\label{sec:ablation}

To highlight our key contributions, we perform a comprehensive ablation study, where a comparison between different configurations show the benefits of our proposed modules, see Tab.~\ref{tab:ablation}. We use REDS for training and evaluation. During training we collect batches composed of $b=32$ samples with crop size $T\times H\times W\times C = 5\times 64\times 64\times 3$. We provide both validation and test set results to emphasize the robustness of our ablation study, but restrict the detailed discussion to REDS (val).

\begin{table}[t]
  \centering
  \resizebox{\textwidth}{!}{
  \begin{tabular}{cccccrr}
    \toprule
    Config.  & \phantom{m}Offsets\phantom{m}  & \phantom{m}Pyramid\phantom{m} & \phantom{m}Attention\phantom{m} & \phantom{m}Features\phantom{m} & REDS4val~\cite{Nah2019reds} (Y) & REDS4~\cite{Nah2019reds} (RGB)  \\
    \midrule
    1&& & & 64 & 28.77/0.7906 & 28.59/0.8155 \\
    2&&  & \checkmark & 64 & 28.95/0.7926 & 28.69/0.8184 \\
    3&\checkmark & & & 64 & 29.82/0.8194 & 29.50/0.8461 \\
    4&\checkmark & \checkmark & & 64 & 30.07/0.8264 & 29.66/0.8507 \\
    5&\checkmark & \checkmark & \checkmark & 64 & \color{blue}30.36/0.8341 & \color{blue}29.97/0.8571 \\
    6&\checkmark & \checkmark & \checkmark & 128 & \color{red}30.77/0.8440  & \color{red}30.49/0.8676  \\
    \bottomrule
  \end{tabular}
  }
  \caption{Ablation study on REDS. All models are trained in the same settings on sequences of 5 frames. \textcolor{red}{Red} denotes best, \textcolor{blue}{blue} denotes second best.}
  \label{tab:ablation}
   \vspace{-0.4cm}
\end{table}

\textbf{Modules}
Configuration 1 is trained without any motion handling and fusion, we only employ the main module $\mathcal{N}$ to propagates a hidden state. The large performance drop compared to all configurations with offsets validates the importance of handling motion and misalignment in VSR. It clearly shows the downsides of naive convolution-based networks for video processing and promotes the necessity for other mechanisms.
Similar conclusions can be drawn from adding our attention-based fusion module $\mathcal{F}$ without providing offsets. Still, a slight gain of $0.18$dB can be achieved by our fusion module even without offsets.
The addition of offsets to configuration 1 significantly improves performance by $1.05$dB in configuration 3. Furthermore, a larger receptive field is attained by application of a pyramid refinement mechanism with simple convolutional fusion instead of our proposed deformable attention. Thus, configuration 4 improves PSNR by $0.25$dB.
Our complete setup with all our proposed modules in combination further boosts performance by $0.29$dB (configuration 5). Moreover, increasing the feature dimension from $64$ to $128$ adds another $0.41$dB. Hence, our proposed DAP network realizes large gains for VSR.
As an aside, in addition to inferior performance of configuration 4, which relies on offset prediction and convolution without our proposed attention mechanism, we observed instabilities during training, which leads us to the conclusion that attention stabilizes training of DNN's in combination with offset prediction.

\begin{table}[h]
  \resizebox{\columnwidth}{!}{
  \centering
  \begin{tabular}{lcccc}
    \toprule
    Configuration\phantom{mmmmm}  &  $\overrightarrow{\phantom{i}\text{DAP-64}\phantom{i}}$ & $\overleftarrow{\phantom{i}\text{DAP-64}\phantom{i}}$& $\overrightarrow{\phantom{i}\text{DAP-128}\phantom{i}}$ & $\overleftarrow{\phantom{i}\text{DAP-128}\phantom{i}}$ \\
    \midrule
    REDS~\cite{Nah2019reds} (RGB) & \phantom{m}29.97/0.8571\phantom{m} & \phantom{m}\textbf{30.16/0.8635}\phantom{m} & \phantom{m}30.49/0.8676\phantom{m} & \phantom{m}\textbf{30.72/0.8751}\phantom{m} \\
    \bottomrule
  \end{tabular}
  }
  \caption{Forward/Reverse ($\rightarrow$/$\leftarrow$) evaluation on REDS4 test set. We evaluate the same model in both directions. 
  }
  \label{tab:reversemode}
  \vspace{-0.4cm}
\end{table}

\textbf{Reverse Evaluation}
A core feature of state-of-the-art bidirectional methods is their ability to fuse information over an entire video offline, which naturally includes aggregation in reverse time order and may have benefits in certain cases. Window-based methods usually aggregate information from future frames with similar potential advantages.
We analyze the effects of relative motion patterns induced by time reversal also in our online setting, since such motion can appear even in non-reversed video, e.g. objects moving away from the camera or a camera that is zooming out.
Therefore, we investigate the difference between forward/backward evaluation, i.e. we evaluate the sequences on the REDS testset in both temporal directions in Tab.~\ref{tab:reversemode}.
Surprisingly, reverse time order aggregation increases performance significantly, i.e. by $+0.19$dB and $+0.23$dB for DAP-64 and DAP-128 respectively. After inspection, we attribute this gain to forward camera motion being more prevalent in these videos. If objects are moving towards the camera, or vice versa, in reverse time order they first appear in higher resolution, simplifying super-resolution for these objects.
Thus, having the opportunity to process video in reverse or having access to future frames, may improve performance for VSR depending on the content, resulting in potential advantages for non-causal methods compared to online algorithms.

\begin{table}[t]
  \centering
  \resizebox{\textwidth}{!}{
  \begin{tabular}{lcccrrrrrrrr}
    \toprule
    & \multirow{2}*{\rotatebox[origin=l]{90}{Unid.}} & \multirow{2}*{\rotatebox[origin=l]{90}{Onl.}} & \multirow{2}*{\rotatebox[origin=l]{90}{R-T.}} & Run & fps & FLOPs & MACs & REDS4\cite{Nah2019reds} & UDM10\cite{PFNL} & Vimeo-90K\cite{xue2019video}\\
    Method &  &  & & [ms] & [1/s] & [G] & [G] & PSNR/SSIM & PSNR/SSIM & PSNR/SSIM \\
    \midrule
     Bicubic & \color{teal}\checkmark &\color{teal}\checkmark&\color{teal}\checkmark & - &- & -&-& 26.14/0.7292 & 28.47/0.8253 & 31.30/0.8687 \\
     TOFlow~\cite{toflow} & \color{teal}\checkmark&\color{red}\xmark&\color{red}\xmark   & - & - &-&-& 27.98/0.7990 & 36.26/0.9438 & 34.62/0.9212\\
     FRVSR~\cite{frvsr} & \color{teal}\checkmark &\color{teal}\checkmark&\color{red}\xmark  & $^*$137 & $^*$7.3 &-&-& - & 37.09/0.9522 & 35.64/0.9319 \\
     DUF~\cite{jo2018duf} & \color{teal}\checkmark &\color{red}\xmark&\color{red}\xmark  & $^*$974 & $^*$1.0 &-&-& 28.63/0.8251 & 38.48/0.9605 & 36.87/0.9447\\
     RBPN~\cite{Haris_2019_CVPR_RBPN} & \color{teal}\checkmark &\color{teal}\checkmark&\color{red}\xmark  & $^*$1507 & $^*$0.7 &-&-& 30.09/0.8590 & 38.66/0.9596 & 37.20/0.9458  \\
     PFNL~\cite{PFNL} & \color{teal}\checkmark &\color{red}\xmark&\color{red}\xmark  & $^*$295 & $^*$3.4 &-&-& 29.63/0.8502 & 38.74/0.9627 & - \\
     MuCAN~\cite{li_mucan} & \color{teal}\checkmark &\color{red}\xmark&\color{red}\xmark& 2'208 & 0.5 & 15'853.2 & 7'922.8 &  \color{blue}30.88/0.8750 & - & -\\
     EDVR-M~\cite{wang2019edvr} & \color{teal}\checkmark  &\color{red}\xmark&\color{red}\xmark & 116 & 8.6 & 925.7 & 462.3 & 30.53/0.8699 & 39.40/0.9663 & 37.33/0.9484  \\
     EDVR~\cite{wang2019edvr} & \color{teal}\checkmark  &\color{red}\xmark&\color{red}\xmark & 348 & 2.9 &4'037.3 & 2'017.3& \color{red}31.09/0.8800 & \color{red}39.89/0.9686 & \color{red}37.81/0.9523 \\
     TGA~\cite{Isobe_2020_CVPR_TGA} &  \color{teal}\checkmark &\color{red}\xmark&\color{red}\xmark& 427 & 2.3 &-&-& -& - & \color{blue}37.59/0.9516\\
     RSDN~\cite{rsdn} & \color{teal}\checkmark  &\color{teal}\checkmark&\color{red}\xmark & 63 & 15.9 &713.2 & 356.3& - & 39.35/0.9653 & 37.23/0.9471\\
     RRN~\cite{isobeRRN} & \color{teal}\checkmark &\color{teal}\checkmark&\color{teal}\checkmark  & \color{red}28 & \color{red}35.7 & \color{blue}387.5 & \color{blue}193.6 & - & 38.96/0.9644 & -\\
     RLSP~\cite{fuoli2019rlsp} & \color{teal}\checkmark &\color{red}\xmark  &\color{teal}\checkmark  & \color{blue}30 & \color{blue}33.3 & 503.7 & 251.8 & - & 38.48/0.9606 & 36.49/0.9403 \\
     DAP-128 (ours) & \color{teal}\checkmark  &\color{teal}\checkmark&\color{teal}\checkmark & 38 & 26.3 & \color{red}330.0 & \color{red}164.8 & 30.59/0.8703 & \color{blue}39.50/0.9664 & 37.29/0.9476\\
     \midrule
     \color{gray}BasicVSR \cite{chan2020basicvsr} & \color{red}\xmark&\color{red}\xmark&\color{red}\xmark   & \color{gray}82 & \color{gray}12.2 &\color{gray}754.3 & \color{gray}376.7& \color{gray}31.42/0.8909 & \color{gray}39.96/0.9694 & \color{gray}37.53/0.9498 \\
     \color{gray}IconVSR \cite{chan2020basicvsr} & \color{red}\xmark &\color{red}\xmark&\color{red}\xmark  & \color{gray}100 & \color{gray} 10.0 &\color{gray}904.9 & \color{gray}451.9& \color{gray}31.67/0.8948 & \color{gray}40.03/0.9694 & \color{gray}37.84/0.9524\\
     \color{gray}BasicVSR++ \cite{chan2021basicvsr++} & \color{red}\xmark &\color{red}\xmark&\color{red}\xmark  & \color{gray} 110 & \color{gray} 9.1 & \color{gray}837.1 & \color{gray}418.1& \color{gray}32.39/0.9069 & \color{gray}40.72/0.9722 & \color{gray}38.21/0.9550\\
    \bottomrule
  \end{tabular}
   }
  \caption{Comparison with state of the art. We compare runtime, frames per second (fps), FLOPs, MACs and PSNR/SSIM metrics on 3 standard benchmarks. Additionally, we denote if a method is unidirectional, i.e. if it can generate output in a single pass (Unid.), can be evaluated strictly online (Onl.), i.e no future frames are needed, and if it can produce video (720p) in real-time (R-T.). All PSNR/SSIM results and runtime measurements marked with * are reported from the respective papers. All other methods are profiled (Run/fps/FLOPs/MACs) in the same settings on a NVIDIA RTX2080Ti by us. \textcolor{red}{Red} denotes best, \textcolor{blue}{blue} denotes second best.}
  \label{tab:sota}
  \vspace{-0.8cm}
\end{table}

\subsection{Comparison with State of the Art}

We compare the performance of our method to state-of-the-art methods on 3 different datasets with diverse properties, see Tab.~\ref{tab:sota}. 
Since we address the \emph{causal} VSR problem, we do not compare to bidirectional methods. Such methods cannot be evaluated in a single pass, which inhibits their application for online processing. Additionally, they have an incomparable advantage as a consequence of access to all frames from a video sequence. However, we still report the results of prominent bidirectional methods~\cite{chan2020basicvsr,chan2021basicvsr++} for reference in Tab.~\ref{tab:sota}.

REDS is a challenging high-resolution ($720\times 1280$) dataset, because large displacements and a non-stabilized camera complicate temporal aggregation. 
UDM10 ($720\times 1272$) on the other hand contains more steady camera motion and continuous movement. 
Vimeo-90K contains short sequences of only $7$ frames for training and testing in small resolution ($256\times 448$). The dataset was released with window-based evaluation in mind, i.e. only the center frame is expected to be restored, which impedes a fair comparison to our recurrent method. To improve comparability we reflect the sequences at the end for 3 frames to compute the metrics on the last frame representing the center frame. Yet, our method still has a disadvantage due to the pressure of estimating each frame up to the end of the sequence, contrary to aggregation from adjacent frames only.

\textbf{REDS}
For comparison with state-of-the-art methods on REDS we extend our training sequences to facilitate learning of longer temporal dependencies. 
We uniformly crop sequences of length $T=15$ with a reduced batch size of $b=8$, as a result of higher memory demand. 
To avoid expensive training from randomly initialized parameters, our model is initialized with the pre-trained weights of configuration 6 from the ablation study and refined for $T=15$.
Training on a larger sequence length $T$ further boosts our performance on REDS by $0.1$dB and $0.0027$ in PSNR and SSIM respectively. With the exception of large and slow models EDVR and MuCAN we significantly surpass all other models in performance with high speed, we even supersede EDVR-M with a reduction in runtime of over $\times 3$ and largely reduced computational demand. DAP-128 impressively handles the complex motion in REDS with only $38$ms per frame. Our method is capable of producing over 24 fps needed for real-time evaluation with the lowest computational complexity among all methods (330.0 GFLOPs/164.8 GMACs).

\textbf{Vimeo-90K/UDM10}
As already mentioned, Vimeo-90K has limits due to its intended evaluation protocol. Nevertheless, DAP-128's performance on Vimeo-90K is comparable with recurrent state-of-the-art method RSDN and window-based EDVR-M, despite EDVR-M's advantage on Vimeo-90K being a window-based method. 
Higher performance is expected for window-based EDVR and TGA as a consequence of larger capacity and aforementioned advantages in evaluation on Vimeo-90K.
On the contrary, UDM10 defines a standard evaluation strategy, more suitable for a realistic and fair comparison. We achieve the second best performance in PSNR with high speed, EDVR is over $9$-times slower with a huge computational complexity overhead. Due to our highly efficient aggregation strategy in our proposed DAP, we significantly surpass recurrent method RSDN both in terms of performance, runtime and computational demand. We largely improve performance over RRN with a gain of $+0.54$dB, lower computational complexity and only slightly increased runtime.

\textbf{Runtime and Computational Complexity}
We set out to design an algorithm to overcome the challenges of online VSR. On top of limitations in temporal information aggregation (only past information available), the fulfilment of hard time constraints and low computational complexity is crucial for this task. With our proposed DAP and overall efficient network design, we achieve the best performance in relation to computational effort and are able to move the Pareto front.
DAP-128 is a high-speed method with the lowest computational demand among all other methods in Tab.~\ref{tab:sota} (330.0 GFLOPs/164.8 GMACs), it is able to reach real-time evaluation speed with over $26$ frames per second and high performance.
Our network design surpasses the fundamental algorithmic design used by EDVR and other window-based methods, since our method achieves better performance overall with much reduced computational complexity and therefore faster runtimes, e.g. EDVR-M vs. DAP-128. 
Note, that EDVR's and other window-based method's runtime measurements do not account for online evaluation due to the opportunity to process windows in parallel offline. Thus, online evaluation most likely leads to higher runtimes in practice.
We also surpass the design of state-of-the-art recurrent networks like RSDN with a significant speed-up and over $2$-times reduced computational demand. 
For an illustration of performance vs. runtime and complexity (number of parameters) please refer to Fig.~\ref{fig:teaser}.

\begin{figure*}[t]
\begin{center}
\includegraphics[width=\linewidth]{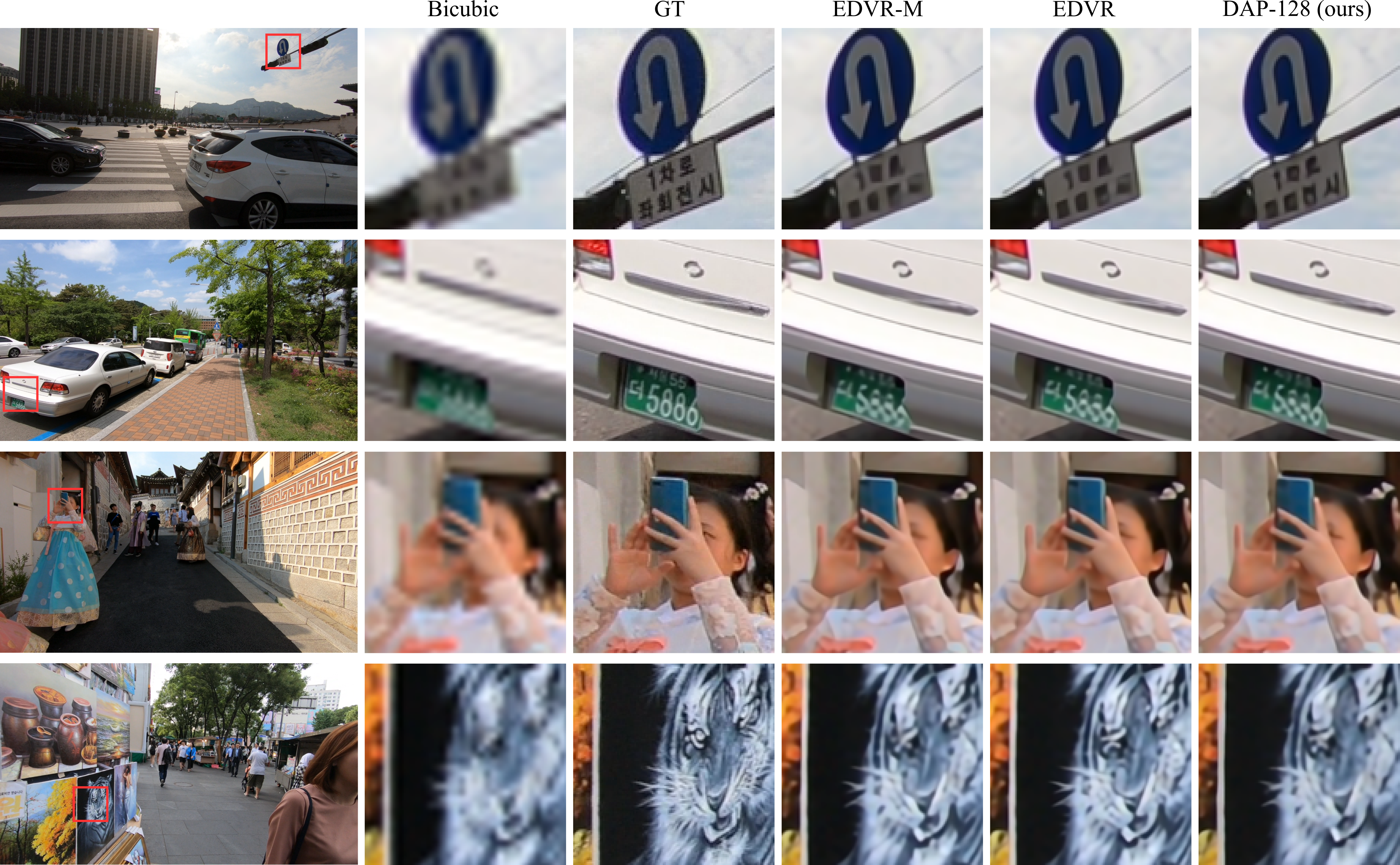}
\end{center}
\vspace{-0.6cm}
   \caption{Visual examples on REDS. Our method achieves competitive performance compared to EDVR-M and EDVR with significantly reduced computational effort.}
\label{fig:visual_examples}
\vspace{-0.85cm}
\end{figure*}

\textbf{Visual Examples}
We provide visual examples for qualitative evaluation in Fig.~\ref{fig:visual_examples}. We compare our method against Bicubic interpolation as a baseline, state-of-the-art method EDVR, its lighter version EDVR-M and the ground truth (GT) on all 4 REDS test sequences. Our method produces high quality frames in accordance to the PSNR/SSIM evaluation in Tab.~\ref{tab:sota}. However, there are individual strengths and weaknesses among all methods. The signals in the first row are restored with higher quality than both EDVR-M and even its heavier version EDVR. On the other hand, the tiger in row 4 is restored in more detail by EDVR. As explained in Sec.~\ref{sec:ablation}, access to future frames can be highly advantageous. EDVR and most other window-based methods access future frames in their window. As a consequence of forward camera motion, the tiger appears in higher resolution in future frames in this particular scene.

\begin{figure}[t]
\begin{center}
\includegraphics[width=\textwidth]{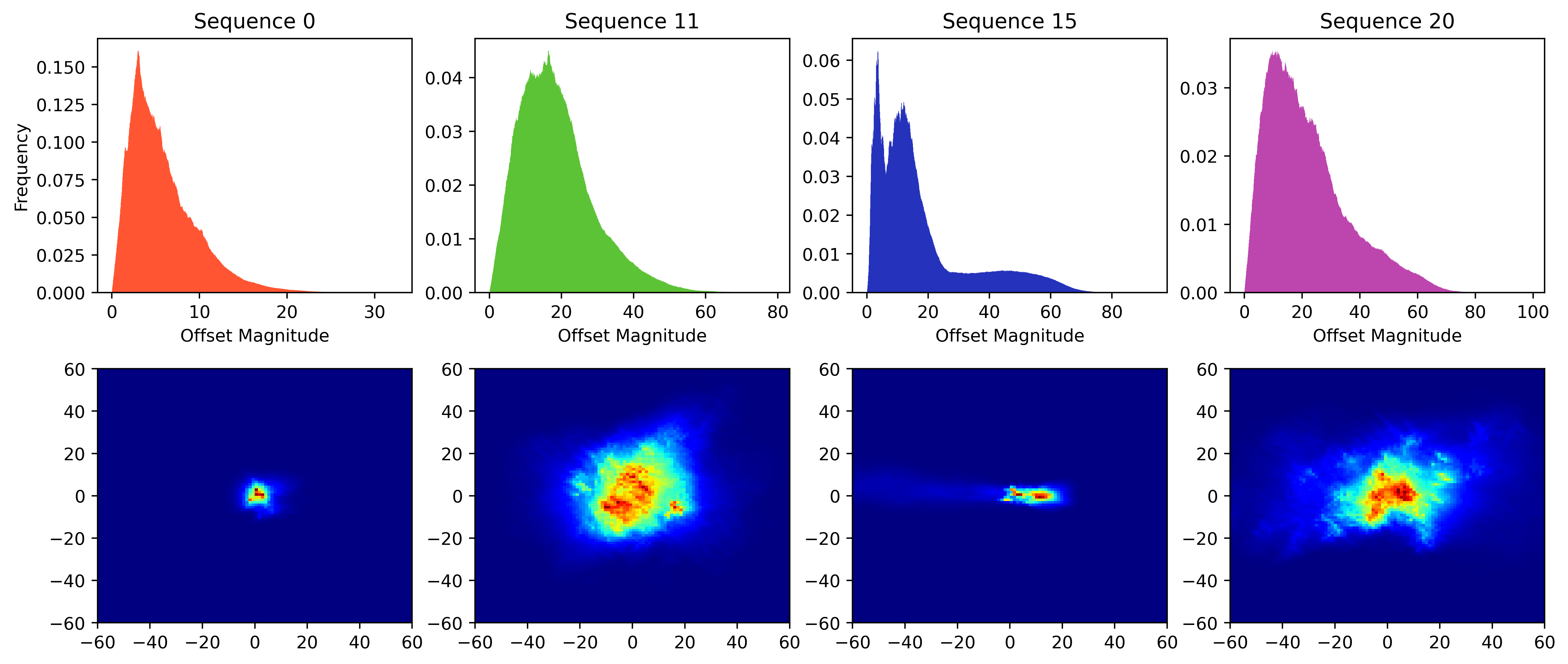}
\end{center}
\vspace{-0.65cm}
   \caption{Analysis of offset locations for DAP-128. Histograms of offset magnitudes are plotted for each sequence in REDS (test set). The bottom row shows corresponding 2D histograms to assess the prominent orientations. Offsets are computed relative to the current frame $x_t$ and are reported in high-resolution domain (in pixels).}
\label{fig:offsets_hist}
\vspace{-0.6cm}
\end{figure}

\textbf{Offset Analysis}
In order to investigate the pixel-dense offsets predicted by our DAP module, we plot the statistics from each video sequence on REDS with 2 types of histograms in Fig.~\ref{fig:offsets_hist}. The top row provides 1D histograms to show the distribution of offset magnitudes in each sequence. 
The offset magnitudes are different, depending on the video content. 
The predicted offsets in sequence 0 are significantly smaller compared to others, which we attribute to more distant objects in the scene after visual inspection. The closer objects appear in a video, the larger their offsets grow.
The second row in Fig.~\ref{fig:offsets_hist} contains 2D histograms, showing the prominent offset directions and magnitudes.
The plots hint at the type of movement in each sequence, e.g. sequence 11 has larger offsets than sequence 0, indicating larger camera movement or more close-up content. Similar arguments can be made about sequence 20. Sequence 15 shows a unique horizontal offset pattern, which is a consequence of a camera pan with horizontally moving objects. There is a concentrated direction of offsets in one direction, and another set of offsets in the opposite direction. These differences are likely caused by opposing movement of background and foreground objects in the scene when the camera pans.

\section{Conclusion}
We address the two main challenges in online VSR; efficient temporal aggregation and misalignment.
Despite the inherent relationship between computational complexity and network capacity, our light-weight designs enables high performance with fast runtimes in the online setting, achieved by our effective attention-based module for combined fusion/alignment of information from the hidden state only.
In contrast to other attention-based solutions for VSR, our proposed DAP avoids exhaustive operations by dynamically attending to the salient locations in the hidden state, thereby significantly reducing the high computational burden associated with attention and transformers. 
Our attention mechanism enables efficient pixel-dense processing, a crucial feature for super-resolution.
Comprehensive experiments and ablation studies reinforce our contributions and provide analysis of our method. We surpass state-of-the art method EDVR-M on two standard benchmarks with a speed-up of over $3\times$ and the lowest computational complexity among all compared methods.

\noindent\textbf{Acknowledgements.}
This work was partly supported by a Huawei Technologies Co. Ltd project, the ETH Z\"urich Fund (OK) and the Humboldt Foundation.

\appendix
\section*{Supplementary Material}

We provide additional visual analysis of the sampling locations predicted by our model, a visual inspection and discussion of reverse evaluation results, analysis of temporal information propagation in our recurrent cell, additional results on Vimeo-90K, further details of our architecture and a qualitative inspection of rendered videos.

\section{Supplementary Videos}

We rendered the results for EDVR-M, EDVR~\cite{wang2019edvr}, DAP-128, ground truth and Bicubic interpolation into videos for visual quality assessment (on REDS). 
An inspection of the video results supports our overall analysis.

\section{Visual Analysis of Sampling Locations}

\begin{figure}[h!]
\begin{center}
\includegraphics[width=\linewidth]{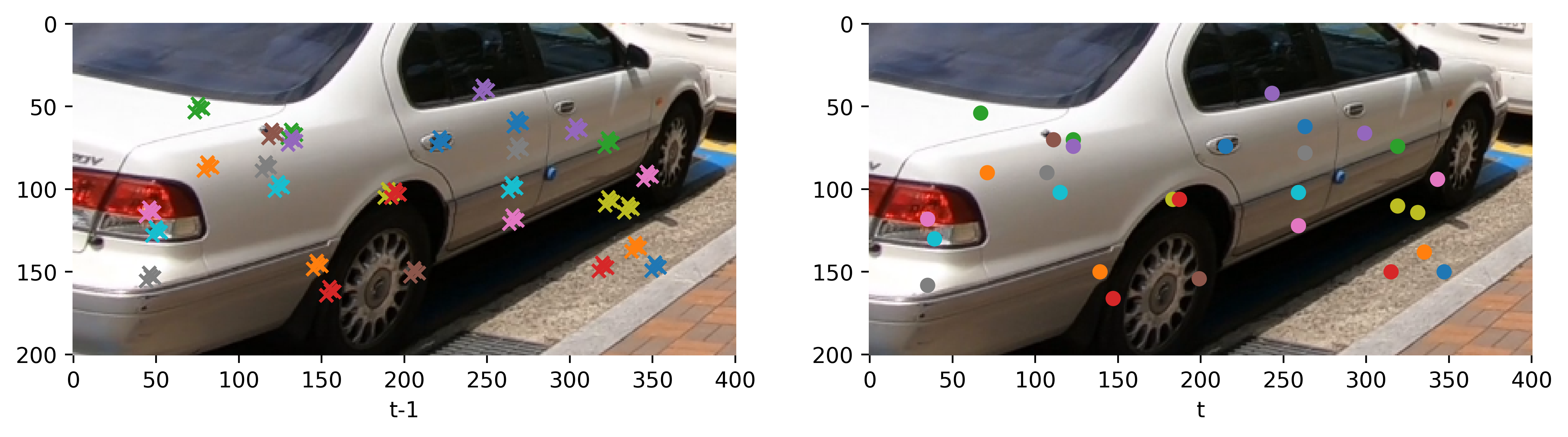}
\end{center}
\vspace{-0.8cm}
   \caption{Illustration of offset predictions. A subset of key/value offset locations in frame $x_{t-1}$ are shown on the left. The corresponding pixel-dense query locations in frame $x_t$ are marked in the same colors on the right. For detailed visual inspection, the offsets are illustrated in the high-resolution domain.}
\label{fig:visual_offsets}
\end{figure}

We analyze the predicted sampling locations of our deformable attention pyramid (DAP) visually in Fig.~\ref{fig:visual_offsets}.
A randomly selected subset of pixel-dense offsets are shown for key/value sampling, overlaid on top of frame $x_{t-1}$. These locations are marked with crosses. Their corresponding pixel-dense query location is shown in frame $x_t$ on the right hand side in Fig.~\ref{fig:visual_offsets}. The points are marked in the same color.
A visual inspection reveals offset predictions that match its corresponding location with high precision. The network learns to attend to offsets that are spread around the point of interest for increased robustness.

\section{Reverse Evaluation - Visual Results}

\begin{figure}[ht!]
\begin{center}
\includegraphics[width=\linewidth]{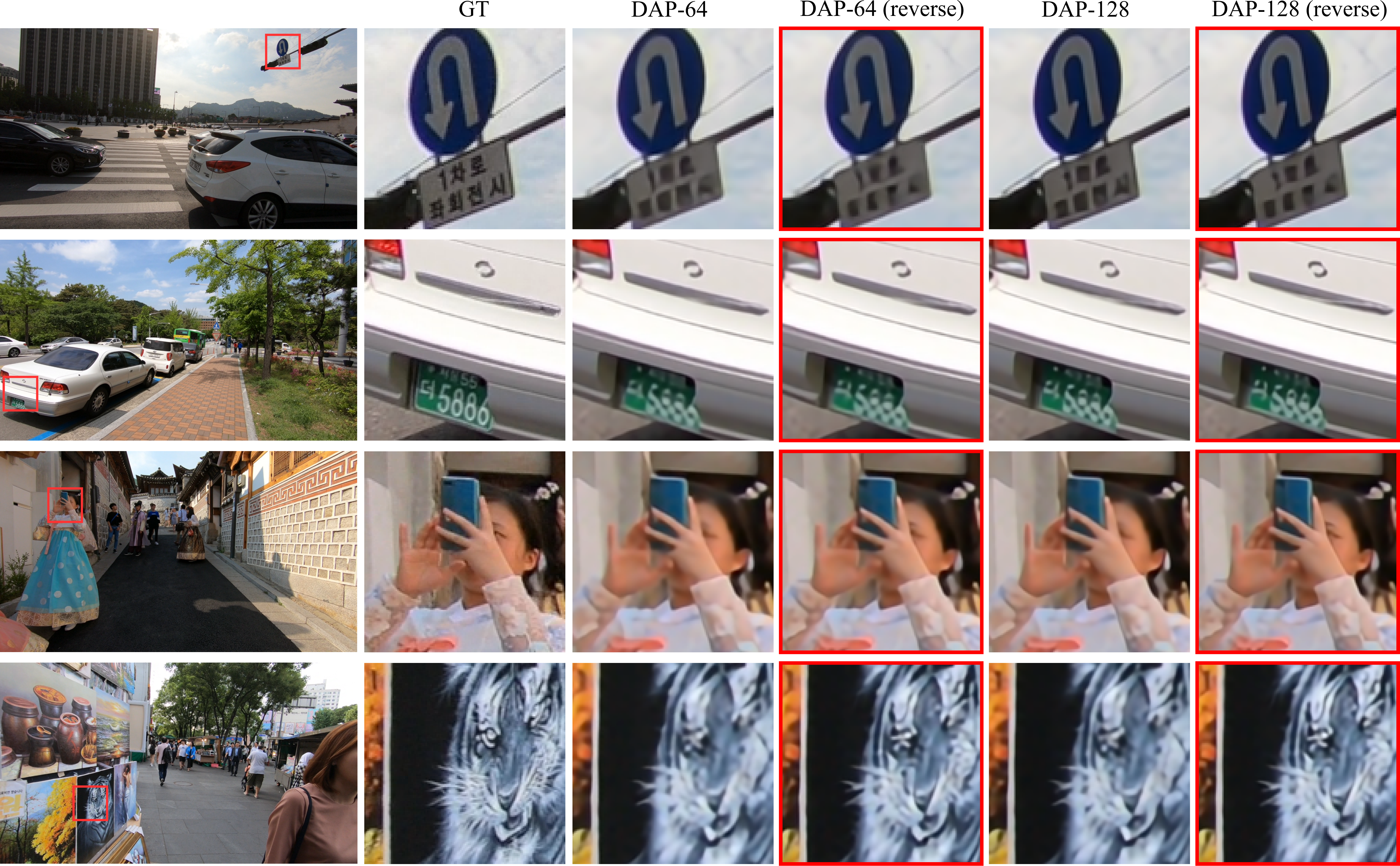}
\end{center}
   \caption{Visual examples on REDS (test set) for forward and reverse mode evaluation. Except for the top row, where the camera motion exhibits opposite behavior, all sequences are better reconstructed in reverse mode. Reverse mode results are highlighted with red borders.}
\label{fig:reversemode}
\end{figure}

We show the qualitative differences between forward and reverse evaluation in Fig.~\ref{fig:reversemode}. The quantitative differences  are investigated in the paper, we list these results again for reference in Tab.~\ref{tab:reversemode_supp}.
The performance gain is attributed to the camera's motion direction as explained in the paper. If an object is first visible in higher resolution (larger), the network can leverage this higher-resolution information about the object in lower resolutions (smaller)
The performance gain is clearly visible in 3 out of 4 sequences from the REDS test set, only the first row reveals better results for forward propagation. The better performing methods include forward camera motion, while the camera in the sequence in the first row pans from left to right. In effect, the sign is first visible in higher resolution (larger) in forward evaluation, leading to better results with the same argument.
The lighter model DAP-64 (reverse) even surpasses the visual quality of DAP-128 in row 4. The tiger's reconstruction shows sharper lines and reveals more details.

\begin{table}[t]
  \resizebox{\columnwidth}{!}{
  \centering
  \begin{tabular}{lcccc}
    \toprule
    Configuration\phantom{mmmmm}  &  $\overrightarrow{\phantom{i}\text{DAP-64}\phantom{i}}$ & $\overleftarrow{\phantom{i}\text{DAP-64}\phantom{i}}$& $\overrightarrow{\phantom{i}\text{DAP-128}\phantom{i}}$ & $\overleftarrow{\phantom{i}\text{DAP-128}\phantom{i}}$ \\
    \midrule
    REDS~\cite{Nah2019reds} (RGB) & \phantom{m}29.97/0.8571\phantom{m} & \phantom{m}\textbf{30.16/0.8635}\phantom{m} & \phantom{m}30.49/0.8676\phantom{m} & \phantom{m}\textbf{30.72/0.8751}\phantom{m} \\
    \bottomrule
  \end{tabular}
  }
  \caption{Forward/Reverse ($\rightarrow$/$\leftarrow$) evaluation on REDS4 test set. We evaluate the same model in both directions. 
  }
  \label{tab:reversemode_supp}
  \vspace{-0.4cm}
\end{table}

\section{Analysis of Temporal Information Propagation}

\begin{figure}[h]
\begin{center}
\includegraphics[width=\linewidth]{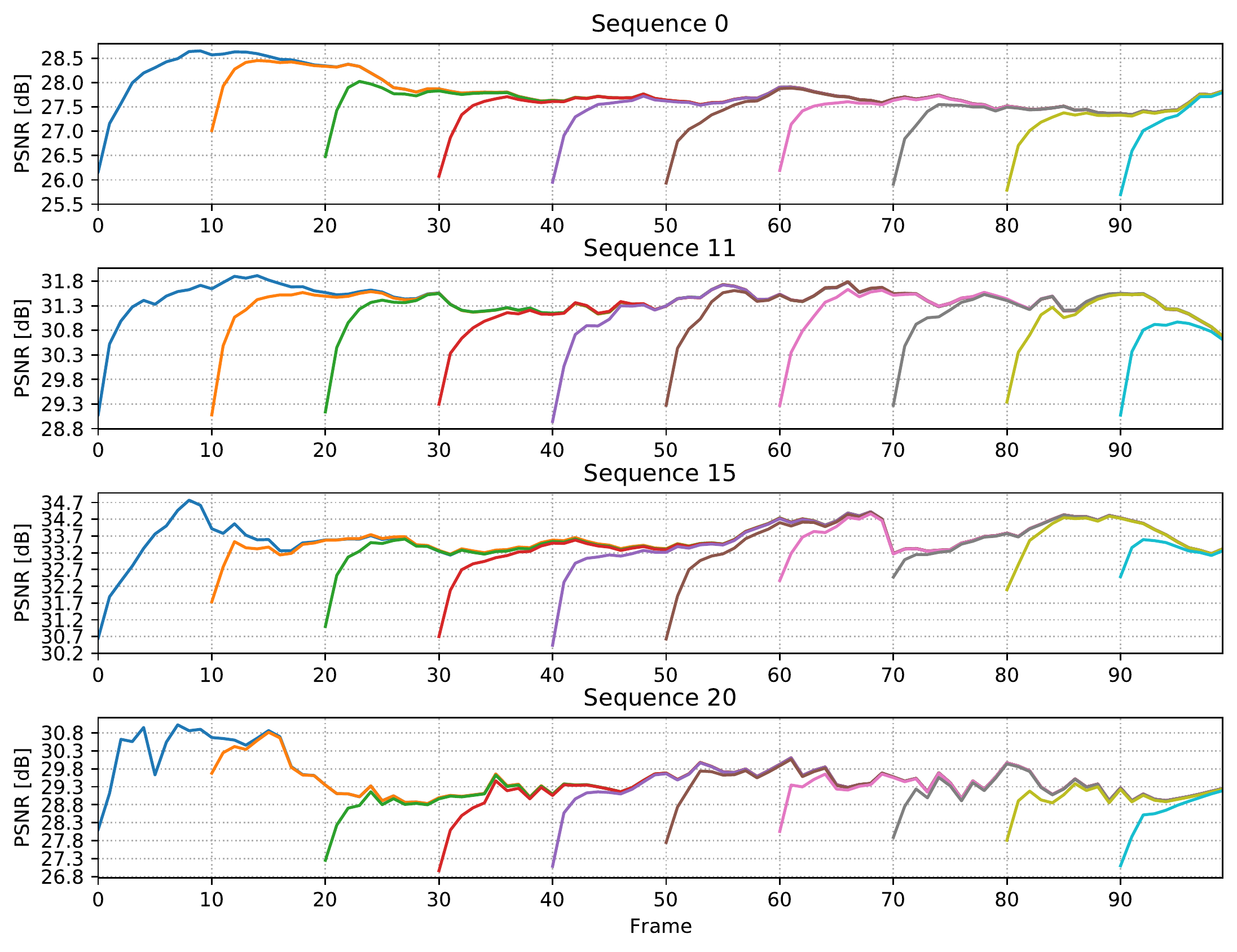}
\end{center}
\vspace{-0.6cm}
   \caption{Analysis of information propagation in our recurrent cell for DAP-128 on REDS (test set). }
\label{fig:temporal_aggregation}
\end{figure}

We investigate the evolution of PSNR in each sequence of REDS (test set)  in Fig.~\ref{fig:temporal_aggregation}. In order to show the importance of temporal aggregation from previous frames, we plot PSNR curves with different starting points, i.e. we initialize an empty hidden state at regular intervals (every 10th frame) and from there evaluate our model until the end of the whole sequence.

We investigate the model DAP-128, which was trained on 15 frames, i.e. the model we use for comparison to state of the art in the paper.
Our proposed recurrent temporal information aggregation mechanism (DAP) efficiently leverages temporal information to improve super-resolution of a single frame. The effect is significant, as shown by the steep initialization curves exposed by subsequent intervals. In some cases - depending on the content - it takes more than 15 frames to reach the previously started model's performance with initial gaps of several dB in PSNR.
Thus, the experiments show the benefits of having access to past information over a long range, efficiently achieved by our DAP aggregation mechanism through the hidden state in our recurrent cell.

\section{Additional Vimeo-90K Results}

We already report full results on REDS and UDM10 - the most relevant datasets due to their high resolution and long sequences - in the main paper along with results for Vimeo-90K with the blur/downsample kernel (BD), which provide a comprehensive overall picture of the compared methods' performance.

For completeness we additionally computed results on Vimeo-90K, obtained by application of Matlab's Bicubic downsampling kernel (BI), see Tab.~\ref{tab:sota_matlab_bi}. We selected the BD setting in the paper as more methods report their results in this setting on Vimeo-90K. The relative performance to the other methods is similar to the BD setting - as generally is the case for different kernels. Thus, the discussion and conclusions in the paper are equally valid after inspection of the BI results. Note, as explained in the paper in Sec.4.2, Vimeo-90K has limitations due to short sequences and its evaluation protocol, which is intended for window-based methods.

\begin{table}[t]
  \centering
  \resizebox{\textwidth}{!}{
  \begin{tabular}{lcccrrrrrrrrr}
    \toprule
    & & & &&&&&&& \multicolumn{2}{c}{Vimeo-90K \cite{xue2019video}}\\
    & \multirow{2}*{\rotatebox[origin=l]{90}{Unid.}} & \multirow{2}*{\rotatebox[origin=l]{90}{Onl.}} & \multirow{2}*{\rotatebox[origin=l]{90}{R-T.}} & Run & fps & FLOPs & MACs & REDS4\cite{Nah2019reds} & UDM10\cite{PFNL} & BD & BI\\
    Method &  &  & & [ms] & [1/s] & [G] & [G] & PSNR/SSIM & PSNR/SSIM & PSNR/SSIM & PSNR/SSIM \\
    \midrule
     Bicubic & \color{teal}\checkmark &\color{teal}\checkmark&\color{teal}\checkmark & - &- & -&-& 26.14/0.7292 & 28.47/0.8253 & 31.30/0.8687 & 31.32/0.8684 \\
     TOFlow~\cite{toflow} & \color{teal}\checkmark&\color{red}\xmark&\color{red}\xmark   & - & - &-&-& 27.98/0.7990 & 36.26/0.9438 & 34.62/0.9212 & 33.08/0.9054\\
     FRVSR~\cite{frvsr} & \color{teal}\checkmark &\color{teal}\checkmark&\color{red}\xmark  & $^*$137 & $^*$7.3 &-&-& - & 37.09/0.9522 & 35.64/0.9319 & -\\
     DUF~\cite{jo2018duf} & \color{teal}\checkmark &\color{red}\xmark&\color{red}\xmark  & $^*$974 & $^*$1.0 &-&-& 28.63/0.8251 & 38.48/0.9605 & 36.87/0.9447 & -\\
     RBPN~\cite{Haris_2019_CVPR_RBPN} & \color{teal}\checkmark &\color{teal}\checkmark&\color{red}\xmark  & $^*$1507 & $^*$0.7 &-&-& 30.09/0.8590 & 38.66/0.9596 & 37.20/0.9458 & 37.07/0.9435  \\
     PFNL~\cite{PFNL} & \color{teal}\checkmark &\color{red}\xmark&\color{red}\xmark  & $^*$295 & $^*$3.4 &-&-& 29.63/0.8502 & 38.74/0.9627 & - & 36.14/0.9363 \\
     MuCAN~\cite{li_mucan} & \color{teal}\checkmark &\color{red}\xmark&\color{red}\xmark& 2'208 & 0.5 & 15'853.2 & 7'922.8 &  \color{blue}30.88/0.8750 & - & - & \color{blue}37.32/0.9465\\
     EDVR-M~\cite{wang2019edvr} & \color{teal}\checkmark  &\color{red}\xmark&\color{red}\xmark & 116 & 8.6 & 925.7 & 462.3 & 30.53/0.8699 & 39.40/0.9663 & 37.33/0.9484 & 37.09/0.9446 \\
     EDVR~\cite{wang2019edvr} & \color{teal}\checkmark  &\color{red}\xmark&\color{red}\xmark & 348 & 2.9 &4'037.3 & 2'017.3& \color{red}31.09/0.8800 & \color{red}39.89/0.9686 & \color{red}37.81/0.9523 & \color{red}37.61/0.9489 \\
     TGA~\cite{Isobe_2020_CVPR_TGA} &  \color{teal}\checkmark &\color{red}\xmark&\color{red}\xmark& 427 & 2.3 &-&-& -& - & \color{blue}37.59/0.9516 & -\\
     RSDN~\cite{rsdn} & \color{teal}\checkmark  &\color{teal}\checkmark&\color{red}\xmark & 63 & 15.9 &713.2 & 356.3& - & 39.35/0.9653 & 37.23/0.9471 & -\\
     RRN~\cite{isobeRRN} & \color{teal}\checkmark &\color{teal}\checkmark&\color{teal}\checkmark  & \color{red}28 & \color{red}35.7 & \color{blue}387.5 & \color{blue}193.6 & - & 38.96/0.9644 & - & -\\
     RLSP~\cite{fuoli2019rlsp} & \color{teal}\checkmark &\color{red}\xmark  &\color{teal}\checkmark  & \color{blue}30 & \color{blue}33.3 & 503.7 & 251.8 & - & 38.48/0.9606 & 36.49/0.9403 & -\\
     DAP-128 (ours) & \color{teal}\checkmark  &\color{teal}\checkmark&\color{teal}\checkmark & 38 & 26.3 & \color{red}330.0 & \color{red}164.8 & 30.59/0.8703 & \color{blue}39.50/0.9664 & 37.29/0.9476 & 37.06/0.9439\\
     \midrule
     \color{gray}BasicVSR \cite{chan2020basicvsr} & \color{red}\xmark&\color{red}\xmark&\color{red}\xmark   & \color{gray}82 & \color{gray}12.2 &\color{gray}754.3 & \color{gray}376.7& \color{gray}31.42/0.8909 & \color{gray}39.96/0.9694 & \color{gray}37.53/0.9498 & \color{gray}37.18/0.9450 \\
     \color{gray}IconVSR \cite{chan2020basicvsr} & \color{red}\xmark &\color{red}\xmark&\color{red}\xmark  & \color{gray}100 & \color{gray} 10.0 &\color{gray}904.9 & \color{gray}451.9& \color{gray}31.67/0.8948 & \color{gray}40.03/0.9694 & \color{gray}37.84/0.9524 & \color{gray}37.47/0.9476 \\
     \color{gray}BasicVSR++ \cite{chan2021basicvsr++} & \color{red}\xmark &\color{red}\xmark&\color{red}\xmark  & \color{gray} 110 & \color{gray} 9.1 & \color{gray}837.1 & \color{gray}418.1& \color{gray}32.39/0.9069 & \color{gray}40.72/0.9722 & \color{gray}38.21/0.9550 & \color{gray}37.79/0.9500\\
    \bottomrule
  \end{tabular}
   }
  \caption{Additional results with Matlab's Bicubic downsampling kernel (BI) on Vimeo-90K. \textcolor{red}{Red} denotes best, \textcolor{blue}{blue} denotes second best.}
  \label{tab:sota_matlab_bi}
  \vspace{-0.8cm}
\end{table}

\section{Method Details}
In this section we present additional details of our modules.
\paragraph{Offset Prediction Block} Our offset prediction network $\mathcal{C}^{l}_\mathcal{S}$ is composed of several light convolution layers with leaky ReLU activations. It features an expansive part followed by a contracting part, all kernels are of size $7\times 7$. We denote the layers as \{$f_{in}, f_{out}$\}, $f_{in}$ represents number of input features, $f_{out}$ stands for number of output features.
Network $\mathcal{C}^{l}_\mathcal{S}$ is a sequence of layers in following configuration:
(\{24, 32\}, \{32,64\},\{64,32\}, \{32,16\}, \{16,8\}). The input consists of $8 + 8 + 4\times 2$ features, 2 input feature maps $f^{l}_t, v^{l}_t$ (encoded features + attention aggregated features) plus the upscaled sampling estimates $U_\uparrow(s^{l+1}_t)$ from the previous level ($k$ locations).
\paragraph{Deformable Attention}
Our proposed attention mechanism consists of 3 processing steps; (1) sampling, (2) encoding and (3) attention.
(1) For each pixel we sample at $k$ locations in $f^l_{t-1}$ according to $s^l \in \mathbb{R}^{H/2^l\times W/2^l\times 2k}$, to obtain $k$ shifted feature vectors. Note, we use $4$ groups to further reduce computations. Each sampled feature vector is encoded into key/value-pairs in step (2), the pixel-dense current frame features $f^l_{t}$ are encoded into query vectors. The feature size for query/key is set to $8$.
Then, cross-attention (3) is performed to aggregate values according to query/key correlations.
In a final step, the hidden state features are aggregated.
To accommodate the larger feature size in the hidden state, we encode its query/key vectors into features of size $8$, but retain the feature size for the values by encoding them in their native dimension (32 per group in DAP-128). This ensures propagation of information in the hidden state without a bottle neck.

\paragraph{Main Processing Block}
The main processing block $\mathcal{N}$ consists of a convolutional layer to aggregate hidden state and input frame $x_t$, followed by 5 repeated fully convolutional IMDN blocks~\cite{Hui-IMDN-2019}. In order to produce the next hidden state $h_t$ and the output $y_t$ we employ another convolutional layer at the end. The input feature dimensions are set to $128+3$ corresponding to the feature size in the hidden state and number of color channels in $x_t$ respectively. Following the repeated IMDN blocks, the final convolution layer produces the high-resolution output $y_t$, represented in low resolution ($48$ features) and the next hidden state $h_t$ ($128$ features for DAP-128).
The high-resolution output frame in RGB is obtained with pixel-shuffle. We also adopt residual learning (nearest neighbor interpolation).

%
%
\bibliographystyle{splncs04}
\bibliography{egbib}
\end{document}